\documentclass[
reprint,
reprint,
showpacs,
amsmath,amssymb,aps,
pra,
]{revtex4-1}
\UseRawInputEncoding
\usepackage{graphicx}
\usepackage{bm}
\usepackage{hyperref}
\usepackage{xcolor}

\begin{document}

\preprint{}

\title{Mode Structure of a Broadband High Gain Parametric Amplifier}

\author{Xin Chen}
\author{Jacob Zhang}
\author{Z. Y. Ou}
\email{zou@iupui.edu}
\affiliation{%
Department of Physics, Indiana University-Purdue University Indianapolis, Indianapolis, IN 46202, USA
}%





\begin{abstract}
High gain parametric amplifier with a single-pass pulsed pump is known to generate broadband twin photon fields that are entangled in amplitude and phase but have complicated spectral correlation. Fortunately, they can be decomposed into independent temporal modes. But the common treatment of parametric interaction Hamiltonian does not consider the issue of time ordering problem of interaction Hamiltonian and thus leads to incorrect conclusion that the mode structure and the temporal mode functions do not change as the gain increases. In this paper, we use an approach that is usually employed for treating nonlinear interferometers and avoids the time ordering issue. This allows us to derive an evolution equation in differential-integral form. Numerical solutions for high gain situation indicate a gain-dependent mode structure that has its mode distributions changed and mode functions broadened as the gain increases.
This study will enable us to have a complete picture of the mode structure of parametric processes and produce high quality quantum sources for a variety of applications of quantum technology.
\end{abstract}


\maketitle

\section{Introduction}

Parametric processes in the high gain regime are the most common and simple processes for generating a variety of quantum states including twin beam states, squeezed states, EPR-entangled states \cite{walls} for the applications in continuous variable quantum information processing, quantum communication, and quantum metrology \cite{reid}. For achieving high gain operation, ultra-short pulse pumping in single-pass configuration \cite{kumar} is usually preferred due to its high concentration of energy and ease of operation. This has been done in optical waveguide structure \cite{wg} and optical fiber systems \cite{guo16} in which spatial modes are well defined. However, due to ultrashort pulses in pumping and dispersion in nonlinear media, the spectral correlation is extremely complicated \cite{guo13}.

Fortunately, the complicated spectral correlation can be decomposed into independent temporal modes \cite{lvo,sil}. This was first pointed out by Law {\it et al.} \cite{law} for the low gain case where two-photon events dominate and used in the analysis of the mode structure for the generation of high quality single- and two-photon states \cite{cui20}. In the high gain regime, the existence of independent pairwise entangled temporal modes was recently confirmed experimentally in a direct measurement of the temporal mode profiles and in subsequent correlation measurement \cite{huo20}.  However, the common treatment of parametric interaction Hamiltonian \cite{sil} only works in the limit of low gain or in the regime of spontaneous emission but fails at high gain because it does not consider the issue of time ordering of the Hamiltonian and thus leads to the incorrect result that temporal modes do not change in the high gain limit \cite{sipe}. Indeed, recent studies with approaches that avoid the time ordering issue showed the spectrum broadening as the gain increases \cite{sha20}. The experiment that directly measured the temporal mode functions also confirmed the change of mode structure and mode functions as the gain increases \cite{huo20}.

The change of mode structure and mode functions with gain is troublesome in the production and applications of high quality quantum sources with quantum entanglement and noise reduction such as EPR entangled states and squeezed states, which require high gain operation in parametric processes. This is because the measurement on these states relies on the homodyne measurement technique in which the mode match between the local oscillator field (LO) and the quantum field is paramount and any mode mismatch is equivalent to losses and introduces extra vacuum noise.
The knowledge of the exact profile of the mode functions will enable us to tailor the shape of LO field to match the quantum field \cite{huo20}. But the change of the mode functions means that we also need to adjust the shape of LO to accommodate the change. The shape of the mode functions is also important for quantum pulse gates \cite{sil11,sil14,ray14} in temporal mode multiplexing.

But so far there is no analysis about how mode structure and mode functions change with the gain.
In this paper, we will investigate the pulse-pumped single-pass parametric processes at arbitrary pumping power. We will use an operator input-output approach that is usually
employed for treating multi-stage nonlinear interferometers \cite{ou12}. This avoids the time ordering issue of the interaction Hamiltonian and allows us to derive a set of coupled operator
evolution equations in differential-integral form. We solve them numerically and analyze the mode structure and mode functions at the final output ports as a function of the pump parameter.

The paper is organized as follows.
In Sec.II, we introduce the input-output approach for the single-mode case of a multi-stage nonlinear interferometer involving parametric processes. In Sec.III, we apply the approach to the evolution of the fields in broadband parametric processes pumped by a pulse for an arbitrary length of nonlinear medium. We discuss mode decomposition in Sec.IV and solve numerically the state evolution in Sec.V to find how mode structure and mode functions change with the gain. We conclude with a discussion in Sec.VI.

\begin{figure}
\centering
\includegraphics[width=4.5cm]{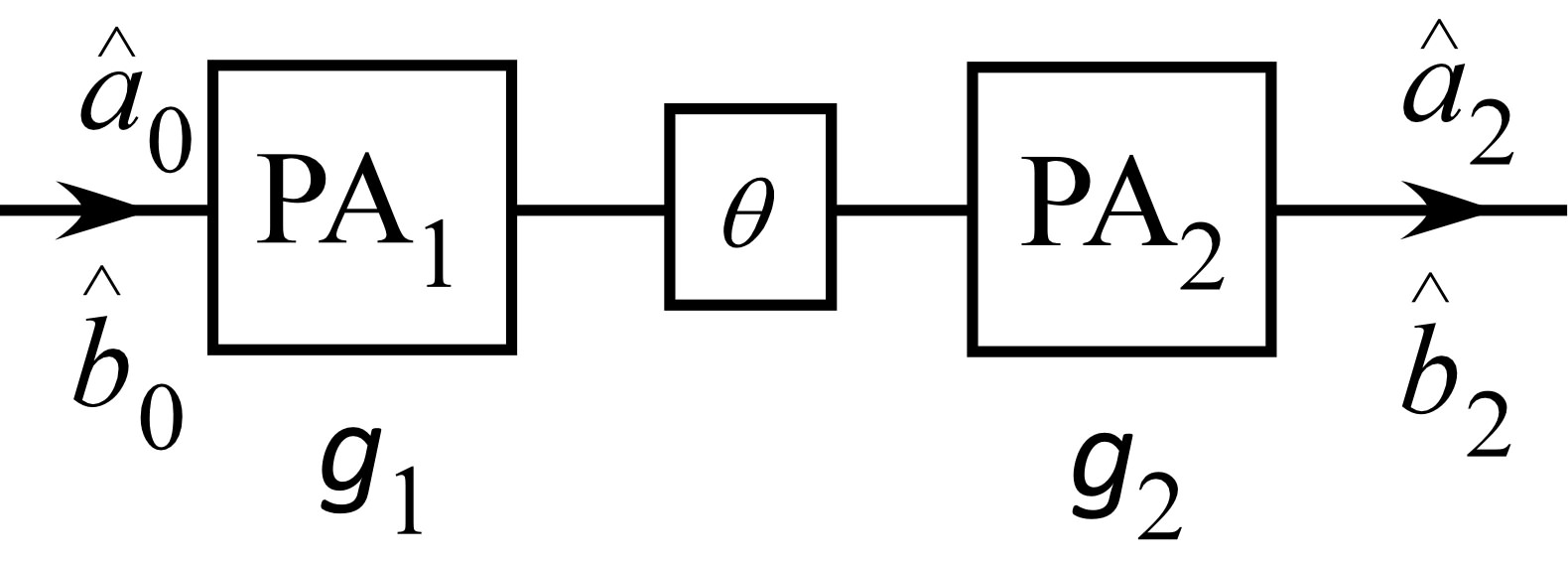}
\caption{An SU(1,1) interferometer with parametric amplifiers (PA$_1$,PA$_2$) in place of beam splitters.
}
\label{fig:SU}
\end{figure}

\section{Multi-stage SU(1,1) interferometers}

In order to reveal the issue of time ordering in the derivation of evolution operator in parametric processes and find ways to tackle it, we consider an SU(1,1) interferometer, shown in Fig.\ref{fig:SU}, which consists of two parametric amplifiers (PA$_1$, PA$_2$) characterized by gain parameters $g_1$, $g_2$ together with a phase shift $\theta$ in between. This interferometer has recently been studied extensively \cite{ouli20} for precision phase measurement beyond standard quantum limit \cite{hud14}, quantum imaging with undetected photons \cite{zei14}, and quantum state engineering \cite{su19,li20}.

The two PAs are described by the Hamiltonians:
\begin{equation}
\label{H-PA}
\hat H_{PA}(\xi_j) = i\hbar \xi_j\hat a^{\dag}\hat b^{\dag} - i\hbar \xi_j^*\hat a\hat b,
\end{equation}
where $j=1,2$.
The input-output relations can be derived from evolution operators $\hat U_j = e^{(1/i\hbar)\hat H_{PA}(\xi_j)t} (j=1,2)$ and are respectively given as
\begin{eqnarray}
\label{in-out}
\hat a_1 &=& G_1\hat a_0 + g_1\hat b_0^{\dag} ,~~ \hat b_1 = G_1\hat b_0 + g_1\hat a_0^{\dag}; \cr
\hat a_2 &=& G_2\hat a_1' + g_2\hat b_1^{'\dag} ,~~ \hat b_2 = G_2\hat b_1' + g_2\hat a_1^{'\dag};
\end{eqnarray}
where $\hat a_1'=\hat a_1 e^{i\theta/2}, \hat b_1'=\hat b_1 e^{i\theta/2}$, the amplitude gains $g_j\equiv (\xi_j/|\xi_j|)\sinh|\xi_jt| (j=1,2)$, and $G_j=\cosh |\xi_jt|$ for interaction time period of $t$. Here we assume the phase shifts are the same for both fields: $\theta_a=\theta_b=\theta/2$.
The outputs of the interferometer are then \cite{ou12}
\begin{eqnarray}
\label{in-out2}
\hat a_2 &=& G_T\hat a_0 + g_T\hat b_0^{\dag},~ \hat b_2 = G_T\hat b_0 + g_T\hat a_0^{\dag},~~~~
\end{eqnarray}
with
\begin{eqnarray}
\label{G-T}
G_T &=& G_1G_2e^{i\theta/2} + g_1^*g_2e^{-i\theta/2} \cr
g_T &=& G_1^*g_2e^{-i\theta/2} + g_1G_2e^{i\theta/2}.
\end{eqnarray}
This shows that we can treat the whole system as one parametric amplifier with equivalent amplitude gains  $g_T$, $G_T$. Furthermore, besides a propagation phase of $e^{i\theta/2}$ for both fields, the extra phase shift $e^{i\theta}$ can be absorbed in $g_2$ by redefining $g_2'\equiv g_2e^{-i\theta}$ or $\xi_2'\equiv \xi_2e^{-i\theta}$. Here $\xi_2'= \xi_2e^{-i\theta}$ takes the propagation phase shift $e^{i\theta}$ into consideration.
Notice that only when $\arg \xi_1= \arg \xi_2-\theta$, other than a common phase of $e^{i\theta/2}$ for both fields, the whole system can be described by an equivalent overall Hamiltonian
$\hat H_T =  \hat H_{PA}(\xi_T) = \hat H_{PA}(\xi_1)+\hat H_{PA}(\xi_2')$ with $\xi_T=\xi_1+\xi_2'$. 
But if $\arg \xi_1 \ne \arg \xi_2-\theta$, then $\hat H_T \ne \hat H_{PA}(\xi_1)+\hat H_{PA}(\xi_2')$. This is because $e^{(1/i\hbar)\hat H_{PA}(\xi_1)t}e^{(1/i\hbar)\hat H_{PA}(\xi_2')t}\ne e^{(1/i\hbar)[\hat H_{PA}(\xi_1)+\hat H_{PA}(\xi_2')]t}$ if $[\hat H_{PA}(\xi_1), \hat H_{PA}(\xi_2')]\ne 0$ when $\arg \xi_1 \ne \arg \xi_2-\theta$.

\begin{figure}
\centering
\includegraphics[width=8.5cm]{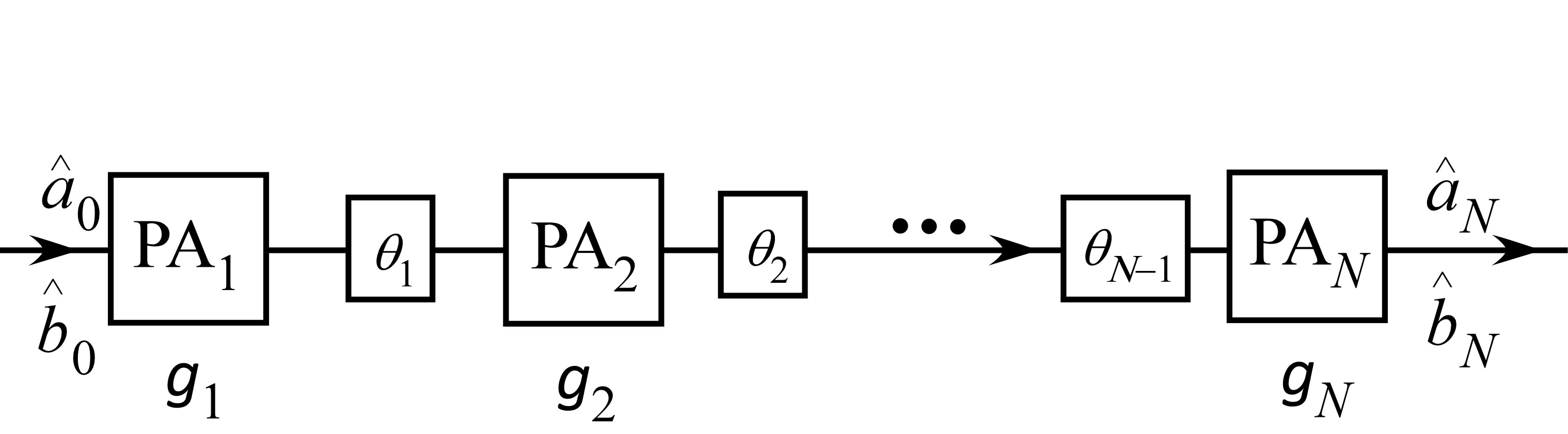}
\caption{A multi-stage SU(1,1)  interferometer.
}
\label{fig:ill}
\end{figure}

This will have an inconvenient consequence when we extend the interferometer to multiple stages \cite{cui20,li20}, as shown in Fig.\ref{fig:ill}. Because the phases in each stage are arbitrary, we therefore cannot write the overall Hamiltonian as the sum of each stage:
\begin{equation}
\label{H-T0}
\hat H_{T} \ne  \sum_j i\hbar(\xi_j'\hat a^{\dag}\hat b^{\dag} - \xi_j^{'*}\hat a\hat b),
\end{equation}
where the phase shifts at each stage are absorbed in the interaction parameter $\xi_j' = \xi_je^{-i\theta_j}$.

To solve this problem, we can proceed  by using repeatedly Eq.(\ref{G-T}) to add each stage and obtain a recursive relation.  Specifically for Eq.(\ref{G-T}), we treat all the stage added up to stage $k$ as the first PA with equivalent amplitude gains $G_T(k), g_T(k)$ and the second PA is the $k+1$-th stage to be added:
\begin{eqnarray}
\label{G-T2}
&&G_T(k+1) = G_T(k)G_{k+1}e^{i\theta_{k}/2} + g_T^*(k)g_{k+1}e^{-i\theta_{k}/2} \cr
&&g_T(k+1) = G_T^*(k)g_{k+1}e^{-i\theta_{k}/2} + g_T(k)G_{k+1}e^{i\theta_{k}/2}. ~~~~
\end{eqnarray}

Unfortunately, we cannot find an analytical expression for the final outputs. In order to have some general idea about the outputs, we consider each stage has an infinitesimally small gain and phase shift whose sizes are proportional to an infinitesimal length scale $\Delta x$ along the field propagation direction: $g_k\approx \zeta(x) \Delta x, \theta_k \approx \eta(x)\Delta x$ and we use location $x=k\Delta x$ to denote the $k$-th stage. When $\Delta x\rightarrow 0$, $G_{k+1} = \sqrt{1+|g_{k+1}|^2} \approx 1 +o(\Delta x)$. So, Eq.(\ref{G-T2}) can be approximated as
\begin{eqnarray}
\label{G-T3}
&&G_T(x+\Delta x) \approx  G_T(x)(1+ i\eta \Delta x/2) + g_T^*(x)\zeta \Delta x  \cr
&&g_T(x+\Delta x) \approx  G_T^*(x)\zeta \Delta x  + g_T(x)(1+ i\eta \Delta x/2). ~~~~
\end{eqnarray}
or
\begin{eqnarray}
\label{G-T4}
&&\frac{d}{dx} G_T(x) =  \zeta(x) g_T^*(x) + \frac{i \eta(x)}{2} G_T(x) \cr
&&\frac{d}{dx} g_T(x) =  \zeta(x) G_T^*(x)  + \frac{i \eta(x)}{2} g_T(x).
\end{eqnarray}
These are the evolution equations for a parametric amplifier with continuous gain function $\zeta(x)$. The phase parameter $\eta(x)$ usually corresponds to phase mismatching. The initial condition is obviously $G_T(0)=1, g_T(0)=0$. It is hard to solve analytically the differential equations if $\zeta(x), \eta(x)$ depend on location $x$. For simplicity, let us assume $\zeta, \eta$ be constant. Then, Eq.(\ref{G-T4}) can be solved analytically and have the following solution:

\noindent If $\zeta \le \eta/2$ and $\eta_0\equiv \sqrt{\eta^2 - 4|\zeta|^2}$, we have
\begin{eqnarray}
\label{G-T5}
&&G_T(x) = \cos \frac{\eta_0 x}{2} + i\frac{\eta}{\eta_0}\sin\frac{\eta_0 x}{2}\cr
&&g_T(x) = \frac{2\zeta}{\eta_0}\sin\frac{\eta_0 x}{2}.
\end{eqnarray}
If $\zeta \ge \eta/2$ and $\zeta_0\equiv \sqrt{|\zeta|^2-(\eta/2)^2}$, we have
\begin{eqnarray}
\label{G-T6}
&&G_T(x) = \cosh (\zeta_0 x) + i\frac{\eta}{2\zeta_0}\sinh(\zeta_0 x)\cr
&&g_T(x) = \frac{\zeta}{\zeta_0}\sinh(\zeta_0 x).
\end{eqnarray}
The exponential growth of the gain when $\zeta \ge \eta/2$ is typical of high gain parametric amplifiers. But the oscillatory low gain behavior when $\zeta \le \eta/2$ is the result of interference as we will see in the following.

In the limit of $\zeta\ll 1$, we have
\begin{eqnarray}
\label{g-T}
G_T(x) &=& e^{i\eta x/2},\cr
g_T(x) &=& \zeta x ~{\rm sinc}(\eta x/2)\equiv  e^{i\eta x/2} x \xi_T \cr
&=& e^{i\eta x/2} \int_0^x dx' \zeta e^{-i\eta x'}.
\end{eqnarray}
The extra phase factor $e^{i\eta x/2}$ extracted out of $g_T$ is for the consistency with $G_T$ and is due to propagation of the fields through the system. The gain parameters in Eq.(\ref{g-T}) are equivalent to an overall Hamiltonian of interaction parameter $\xi_T\equiv  \zeta  e^{-i\eta x/2}{\rm sinc}(\eta x/2)$ with evolution time $t$ replaced by $x$:
\begin{equation}
\label{H-T}
\hat H_{T} = i\hbar(\xi_T\hat a^{\dag}\hat b^{\dag} - \xi_T^*\hat a\hat b),
\end{equation}
which can thought of as the sum of all the stages:
\begin{eqnarray}
\label{H-T2}
\hat H_{T} &=& \frac{1}{x}\int_0^x dx'  i\hbar \zeta e^{-i\eta x'}\hat a^{\dag}\hat b^{\dag} +h.c. \cr
&=&  \frac{1}{x} \int_0^x \hat H(d\xi').
\end{eqnarray}
Here, $d\xi' = \zeta e^{-i\eta x'} dx'$ is the infinitesimal gain parameter for each infinitesimal stage. Note that this equivalence is true only when $\zeta(x)= $ constant and $\eta(x)= $ constant.

As a matter of fact, when gain parameters $|g_1|, |g_2| \ll 1$, we can add the two Hamiltonian to obtain overall Hamiltonian: $\hat H_T =  \hat H_{PA}(\xi_T) = \hat H_{PA}(\xi_1)+\hat H_{PA}(\xi_2')$. This can be seen from evolution operator
\begin{eqnarray}
\label{U-T}
\hat U_{T} &=&  \hat U_2(\xi_2) \hat U_1(\xi_1') = e^{\hat H(\xi_2)\Delta x/i\hbar}e^{\hat H(\xi_1')\Delta x/i\hbar} \cr
&\approx & [1+\hat H(\xi_2)\Delta x/i\hbar][1+\hat H(\xi_1')\Delta x/i\hbar]\cr
&\approx & 1+\hat H(\xi_2)\Delta x/i\hbar+ \hat H(\xi_1')\Delta x/i\hbar \cr
&= & 1 + \hat H_T\Delta x/i\hbar \approx  e^{\hat H_T\Delta x/i\hbar},
\end{eqnarray}
Here, we assumed $|g_1|=|\xi_1|\Delta x\ll 1, |g_2|= |\xi_1'|\Delta x \ll 1$ and $\xi_1'=\xi_1e^{-i\theta}$ with phase shift $e^{-i\theta}$ included. So, when the amplitude gains $|g_j|\ll 1$, we can simply add the Hamiltonian of each stage:
\begin{eqnarray}
\label{H-T3}
\hat H_{T} &=&  \sum_j i\hbar(\xi_j'\hat a^{\dag}\hat b^{\dag} - \xi_j^{'*}\hat a\hat b)
\end{eqnarray}
and
\begin{eqnarray}
\label{g-T3}
g_{T} &=&  \sum_k g_k' = \sum_k g_k e^{-i\theta_k}.
\end{eqnarray}
The above can also be thought of as a result of two-photon interference among the pair of photons generated by each stage \cite{cui20,li20}. This can be confirmed by looking at the output state for vacuum input:
\begin{eqnarray}
\label{Psi-T}
|\Psi\rangle_{T} &=&  \hat U_T |0\rangle \cr
&\approx & |0\rangle + (\sum_k g_k')|1_a,1_b\rangle\cr
&=& |0\rangle + \sum_k |\Psi\rangle_k
\end{eqnarray}
with $|\Psi\rangle_k$ as the two-photon state generated by $k$-th stage.

Notice that Eq.(\ref{H-T3}) is true only if the overall amplitude gain $g_T$ is much smaller than 1 so that the last step of Eq.(\ref{U-T}) stands. For the high gain case, Eq.(\ref{H-T3}) does not stand and we have to resort to Eq.(\ref{G-T2}) or Eq.(\ref{G-T6}). This will pose serious problem in finding solution for a broad band parametric amplifier in the high gain regime.

\section{Broadband Parametric Amplifier}

When parametric processes are pumped by high power pulses, broadband parametric amplification is achieved. They can be used to produce quantum entangled fields with a wide bandwidth. The traditional treatment of this situation is to start with a multi-mode Hamiltonian of the form \cite{sil,guo13}
\begin{eqnarray}
\hat H_M  &=&  \chi \int d\omega_1d\omega_2 d\omega_3 \Psi(\omega_1,\omega_2,\omega_3)\hat a^{\dag}(\omega_1) \hat b^{\dag}(\omega_2)\cr
 &&\hskip 0.3in \times A_p(\omega_3)e^{i(\omega_1+\omega_2-\omega_3)t}+ h.c. , ~~~~\label{H-M}
\end{eqnarray}
where subscript ``M" denotes multi-mode, $\chi$ is some parameter proportional to the nonlinear coefficient of nonlinear medium of length $L_0$, $A_p(\omega_3)$ is the spectral amplitude of the pump field, and $\Psi(\omega_1,\omega_2,\omega_3)$ is obtained from spatial integration:
\begin{equation}
\Psi(\omega_1,\omega_2,\omega_3) \equiv \int_0^{L_0} dz e^{-iz \Delta k} = { L_0 } \frac {\sin \beta}{ \beta}e^{-i\beta} \label{phi}
\end{equation}
with $\beta \equiv \Delta k L_0/2$ and $\Delta k \equiv k_1+k_2-k_3$ as the phase mismatching.
We then find evolution operator as
\begin{eqnarray}\label{U}
\hat U = \exp\left\{\frac{1}{ i\hbar}\int_{-\infty}^{\infty} dt \hat H_M\right\}
\end{eqnarray}
where time integration gives rise to a delta-function $\delta(\omega_1+\omega_2-\omega_3) $ and the integrated Hamiltonian has the form of
\begin{eqnarray}\label{H}
\int dt\hat H_M   = i\hbar G \int d \omega_1 d\omega_2\Phi(\omega_1,\omega_2)\hat a ^{\dag}(\omega_1)\hat b^{\dag}(\omega_2)+ h.c.,\cr &&
\end{eqnarray}
with $G\equiv \chi /C$ as a dimensionless gain parameter such that
\begin{equation}
\Phi(\omega_1,\omega_2) \equiv {2\pi C L_0 } \frac{\sin \beta}{ \beta}e^{-i\beta} A_p(\omega_1+\omega_2) \label{Phi}
\end{equation}
is normalized: $\int d \omega_1 d\omega_2|\Phi(\omega_1,\omega_2)|^2 =1$.
In general, this gives rise to a complicated coupling of different frequency components at the outputs:
\begin{eqnarray}
\label{ha}
&&\hat{a}^{(o)}(\omega_1)=\hat{U}^{\dag} \hat{a}(\omega_1) \hat{U}\cr &&\hskip 0.15in =\int h_{1a}(\omega_1,\omega_1')\hat{a}(\omega_1') d\omega_1'+ \int h_{2a}(\omega_1,\omega_2')\hat{b}^{\dag}(\omega_2') d\omega_2'\cr &&
\end{eqnarray}
\begin{eqnarray}
\label{hb}
&&\hat{b}^{(o)}(\omega_2)=\hat{U}^{\dag} \hat{b}(\omega_2) \hat{U}\cr &&\hskip 0.15in=\int h_{1b}(\omega_2,\omega_2')\hat{b}(\omega_2') d\omega_2'+ \int h_{2b}(\omega_1', \omega_2)\hat{a}^{\dag}(\omega_1') d\omega_1' ,\cr &&
\end{eqnarray}
where $\hat{a}^{(o)},\hat{b}^{(o)}$ are the outputs at the end of nonlinear medium and $\hat{a}(\omega_1),\hat{b}(\omega_2)$ are those at the start. But we can make a singular value decomposition (SVD) of the joint spectrum function (JSF) $\Phi(\omega_1,\omega_2)$:
\begin{equation}
\Phi(\omega_1,\omega_2) =\sum_k r_k \psi_k(\omega_1)\varphi_k(\omega_2), \label{SDV}
\end{equation}
where $\{\psi_k, \phi_k\}$ are two sets of ortho-normal functions: $\int d\omega_1 \psi_j^*(\omega_1)\psi_k(\omega_1)=\delta_{jk} = \int d\omega_2 \varphi_j^*(\omega_2)\varphi_k(\omega_2)$ and $\{r_k\}$ are non-negative numbers satisfying $\sum_k r^2_k=1$. Then
Eq.(\ref{H}) becomes
\begin{eqnarray}\label{H2}
\int dt\hat H_M   = i\hbar G \sum_k r_k \hat A_k^{\dag}\hat B_k^{\dag}+ h.c.
\end{eqnarray}
with $\hat A_k\equiv \int d\omega_1 \psi_j^*(\omega_1)\hat a(\omega_1), \hat B_k \equiv \int d\omega_2 \varphi_j^*(\omega_2)\hat b(\omega_2)$ satisfying $[\hat A_j,\hat A_k^{\dag}]=\delta_{jk}=[\hat B_j,\hat B_k^{\dag}]$. Together with Eq.(\ref{U}),
this leads to de-coupling of the different temporal modes $\hat A_k$ and $\hat B_k$ for the two output fields:
\begin{eqnarray}
\hat{A}_{k}^{(o)}&=& \cosh {(r_k G)} \hat{A}_{k} + \sinh {(r_k G)}\hat{B} ^\dagger _{k},\cr
\hat{B}_{k}^{(o)} &=& \cosh {(r_k G)} \hat{B}_{k} + \sinh {(r_k G)}\hat{A} ^\dagger_{k}, \label{Ak-evl}
\end{eqnarray}

Unfortunately, it was pointed out \cite{sipe} that the evolution operator in Eq.(\ref{U}) is not correct for the Hamiltonian in Eq.(\ref{H-M}) because $[\hat H_M(t),\hat H_M(t')] \ne 0$ for $t\ne t'$. The reason is the same as those for Eq.(\ref{H-PA}). But Eqs.(\ref{ha},\ref{hb}) are still correct for the Hamiltonian in Eq.(\ref{H-M}).

\begin{figure}
\centering
\includegraphics[width=5.5cm]{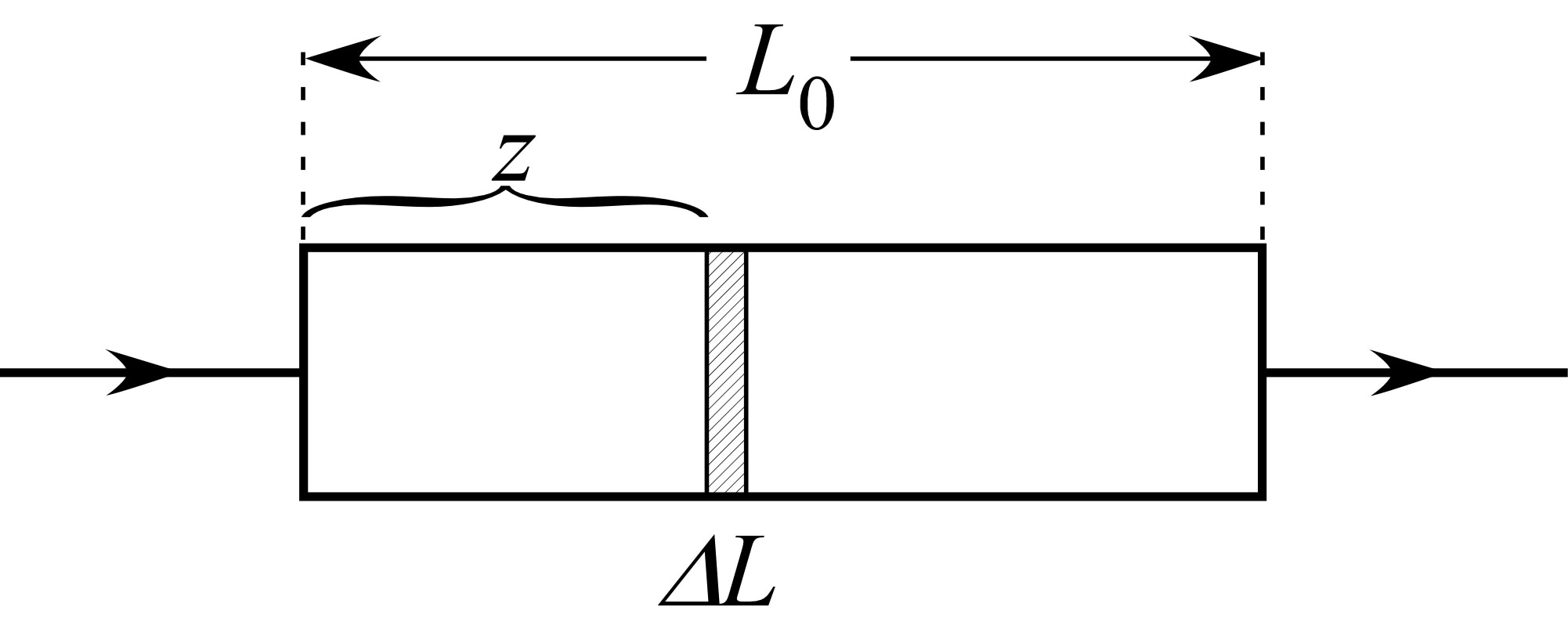}
\caption{A parametric amplifier from a single-pass pulse-pumped nonlinear medium.
}
\label{NM}
\end{figure}

In order to treat this in a correct manner, we can apply the same approach in previous section. Consider a nonlinear medium of length $L$ which is divided into a small segment of size $\Delta L$, as shown in Fig.\ref{NM}. Let us treat the small segment first. The Hamiltonian is given from Eq.(\ref{H-M}) for the small segment as
\begin{eqnarray}
\hat H(z,\Delta L)  &=&  \chi \int d\omega_1d\omega_2 d\omega_3 \int_z^{z+\Delta L} dz' e^{-iz'\Delta k} \hat a^{\dag}(\omega_1,z) \cr
 &&\hskip 0.1in \times  \hat b^{\dag}(\omega_2,z) A_p(\omega_3)e^{i(\omega_1+\omega_2-\omega_3)t}+ h.c. , ~~~~~~\label{H-M2}
\end{eqnarray}
where the spatial integration starts at $z$ instead of $0$ because the $\Delta L$ segment is located at $z$ inside the medium and we assume that there is no pump depletion. For $\Delta L = dL \rightarrow 0$, we have
\begin{eqnarray}
\hat H(z,dL)  &=&   \chi dL \int d\omega_1d\omega_2 d\omega_3 e^{-iz\Delta k} \hat a^{\dag}(\omega_1,z)\hat b^{\dag}(\omega_2,z)  \cr
 &&\hskip 0.1in \times  A_p(\omega_3)e^{i(\omega_1+\omega_2-\omega_3)t}+ h.c. + o(dL).~~~~~~~~\label{H-M3}
\end{eqnarray}
The evolution operator for this segment is given by the Dyson series:
\begin{eqnarray}
\hat U(z,\Delta L)  =  1 + \sum_{n=1} \hat U_n \label{Uz}
\end{eqnarray}
with
\begin{eqnarray}
\hat U_n &=& \left(\frac{1}{i\hbar}\right)^n \int_{-\infty}^{\infty} dt_1 \int_{-\infty}^{t_1}dt_2...\int_{-\infty}^{t_{n-1}}dt_n \cr && \hskip 0.3in \times \hat H(z,dL,t_1)...\hat H(z,dL,t_n).\label{Un}
\end{eqnarray}
Eq.(\ref{Uz}) becomes Eq.(\ref{U}) if $[\hat H(z,dL,t_1),\hat H(z,dL,t_2)]=0$ but it is not true for $\hat H(z,dL)$ in Eq.(\ref{H-M3}). On the other hand, for $dL\rightarrow 0$, we have
\begin{eqnarray}
\hat U(z,d L)  =  1 +  \hat U_1 +o(dL)\approx 1+ \int dt \hat H(z,dL).~~~~ \label{Uz2}
\end{eqnarray}
So, we obtain the evolution of the field operators:
\begin{eqnarray}
\hat a(\omega_1, z+d L) &=& \hat U^{\dag}(z,d L) \hat a(\omega_1, z) \hat U(z,d L) \cr &=&  \hat a(\omega_1, z) + [\hat a(\omega_1,z), \int dt \hat H(z,dL)]\cr
&=& \hat a(\omega_1, z) + 2\pi\chi dL \int d\omega_2  \hat b^{\dag}(\omega_2,z) \cr
 &&\hskip 0.3in \times  e^{-iz\Delta k} A_p(\omega_1+\omega_2),~~~~ \label{a-o}
\end{eqnarray}
where the time integral gives rise to a $\delta$-function for $\omega_1+\omega_2-\omega_3$ and $\Delta k = \Delta k_{|\omega_3=\omega_1+\omega_2}$.
With $d\hat a(\omega_1, z)/dz$  $= [\hat a(\omega_1, z+d L) -\hat a(\omega_1, z)]/dL$, we have
\begin{eqnarray}
&&\frac{d}{dz}\hat a(\omega_1, z)
= 2\pi\chi \int d\omega_2   e^{-iz\Delta k} A_p(\omega_1+\omega_2) \hat b^{\dag}(\omega_2,z).\cr && \label{dadz}
\end{eqnarray}
Likewise, we obtain
\begin{eqnarray}
&&\frac{d}{dz}\hat b(\omega_2, z)
= 2\pi\chi \int d\omega_1   e^{-iz\Delta k} A_p(\omega_1+\omega_2) \hat a^{\dag}(\omega_1,z).\cr && \label{dbdz}
\end{eqnarray}
Note that Eqs.(\ref{dadz},\ref{dbdz}) are in a similar form as those derived in Refs.\citenum{sha20},\citenum{que20} with different methods. Using the input-output relation in Eq.(\ref{ha}) at location $z$, we obtain
\begin{eqnarray}
\frac{d}{dz}h_{1a}(\omega_1, \omega_1',z)
&=&  \int d\omega_2  f(\omega_1, \omega_2)  h_{2b}^*(\omega_1',\omega_2,z)\cr
\frac{d}{dz}h_{2b}(\omega_1', \omega_2,z)
&=&  \int d\omega_1  f(\omega_1, \omega_2)  h_{1a}^*(\omega_1,\omega_1',z),~~~~~~ \label{h1a2b}
\end{eqnarray}
and
\begin{eqnarray}
\frac{d}{dz}h_{1b}(\omega_2, \omega_2',z)
&=&  \int d\omega_1  f(\omega_1, \omega_2)  h_{2a}^*(\omega_1,\omega_2',z)\cr
\frac{d}{dz}h_{2a}(\omega_1, \omega_2',z)
&=&  \int d\omega_2  f(\omega_1, \omega_2)  h_{1b}^*(\omega_2,\omega_2',z),~~~~~~ \label{h1b2a}
\end{eqnarray}
where $f(\omega_1, \omega_2) \equiv 2\pi\chi e^{-iz\Delta k} A_p(\omega_1+\omega_2)$. Since we have $\hat a^{(o)}(\omega_1, z=0) = \hat a(\omega_1)$ and $\hat b^{(o)}(\omega_2, z=0) = \hat b(\omega_2)$ at the start,  the initial condition is
\begin{eqnarray}
h_{1a}(\omega_1, \omega_1', z=0) &=& \delta(\omega_1-\omega_1')\cr
h_{1b}(\omega_2, \omega_2', z=0) &=& \delta(\omega_2-\omega_2')\cr
h_{2a}(\omega_1, \omega_2', z=0) &=& 0\cr
h_{2b}(\omega_2, \omega_1', z=0) &=& 0. \label{ini}
\end{eqnarray}
From Eqs.(\ref{h1a2b},\ref{h1b2a}) and initial conditions in Eq.(\ref{ini}), we can verify that
\begin{eqnarray}\label{h-rel}
&&\int d\omega_1'h_{1a}(\omega_1,\omega_1',z)h_{1a}^*(\bar\omega_1,\omega_1',z)\cr
&& ~~- \int d\omega_2'h_{2a}(\omega_1,\omega_2',z)h_{2a}^*(\bar\omega_1,\omega_2',z) = \delta(\omega_1-\bar\omega_1),\cr
&&\int d\omega_2'h_{1b}(\omega_2,\omega_2',z)h_{1b}^*(\bar\omega_2,\omega_2',z)\cr
&& ~~- \int d\omega_1'h_{2b}(\omega_1',\omega_2,z)h_{2b}^*(\omega_1',\bar\omega_2,z) = \delta(\omega_2-\bar\omega_2),\cr
&&\int d\omega_1'h_{1a}(\omega_1,\omega_1',z)h_{2b}(\omega_1', \bar\omega_2,z)\cr
&& ~~- \int d\omega_2'h_{1b}(\bar\omega_2,\omega_2',z)h_{2a}(\omega_1,\omega_2',z) = 0, \label{dh12}
\end{eqnarray}
These relations guarantee the commutation relations $[\hat a^{(o)}(\omega_1), \hat a^{(o)\dag}(\bar\omega_1)] =\delta(\omega_1-\bar\omega_1)$, $[\hat b^{(o)}(\omega_2),$ $\hat b^{(o)\dag}(\bar\omega_2)]$ $=\delta(\omega_2-\bar\omega_2)$, $[\hat a^{(o)}(\omega_1), \hat b^{(o)}(\bar\omega_2)] =0$ from Eq.(\ref{ha}).

\section{Eigen-modes of high gain parametric processes}
Although $h$-functions have very complicated form, we can in general use singular value decomposition method to decompose them as
\begin{eqnarray}
&&h_{1a}(\omega,\omega') = \sum_{k}r_{1a}^{(k)}\psi_{1a}^{(k)}(\omega)\phi_{1a}^{(k)}(\omega') \cr
&&h_{2a}(\omega,\omega') =\sum_{k}r_{2a}^{(k)}\psi_{2a}^{(k)}(\omega)\phi_{2a}^{(k)}(\omega') \cr
&&h_{1b}(\omega,\omega') =\sum_{k}r_{1b}^{(k)}\psi_{1b}^{(k)}(\omega)\phi_{1b}^{(k)}(\omega') \cr
&&h_{2b}(\omega,\omega') = \sum_{k}r_{2b}^{(k)}\psi_{2b}^{(k)}(\omega)\phi_{2b}^{(k)}(\omega').
\end{eqnarray}
Because of relations in Eq.(\ref{dh12}), it can be shown (see Appendix) that $\psi_{1a}^{(k)}(\omega) = \psi_{2a}^{(k)}(\omega) \equiv \psi_{k}^{(a)}(\omega),~\psi_{1b}^{(k)}(\omega) = $ $ \psi_{2b}^{(k)}(\omega) \equiv \psi_{k}^{(b)}(\omega)$, $\phi_{1a}^{(k)}(\omega')= \phi_{2b}^{(k)*}(\omega') \equiv \phi_{k}^{(a)}(\omega')$, $\phi_{1b}^{(k)}(\omega')$ $= \phi_{2a}^{(k)*}(\omega') \equiv \phi_{k}^{(b)}(\omega')$, and $r_{1a}^{(k)}=r_{1b}^{(k)}  \equiv \cosh r_k G$, $r_{2a}^{(k)}=r_{2b}^{(k)}=\sqrt{r_{1a}^{(k)2} -1}= \sinh r_k G$.~Here, $\psi_k^{(a)}(\omega), \phi_k^{(a)}(\omega), \psi_k^{(b)}(\omega), \phi_k^{(b)}(\omega)$ are four sets of ortho-normal mode functions satisfying $\int d\omega \psi_k^{(a,b)*}(\omega)$ $\psi_{k'}^{(a,b)} (\omega) = \delta_{kk'}$, $\int d\omega \phi_k^{(a,b)*}(\omega)\phi_{k'}^{(a,b)} (\omega) = \delta_{kk'}$ and $r_k$'s are normalized mode coefficients satisfying $\sum_kr_k^2=1$ with $G$ being some parameter depending on $\chi$. So, $h$-functions are in the form of
\begin{eqnarray}\label{hrk}
&&h_{1a}(\omega_1,\omega_1',z) = \sum_k \cosh(r_kG) \psi_k^{(a)}(\omega_1)\phi_k^{(a)}(\omega_1'),\cr &&h_{2a}(\omega_1,\omega_2',z) = \sum_k \sinh(r_kG) \psi_k^{(a)}(\omega_1)\phi_k^{(b)*}(\omega_2'), \cr
&&h_{1b}(\omega_2,\omega_2',z)= \sum_k \cosh(r_kG) \psi_k^{(b)}(\omega_2)\phi_k^{(b)}(\omega_2'),\cr &&h_{2b}(\omega_1,\omega_2',z) = \sum_k \sinh(r_kG) \psi_k^{(b)}(\omega_1)\phi_k^{(a)*}(\omega_2').~~~~
\end{eqnarray}
With these relations and orthonormal relations for $\psi_k^{(a,b)}(\omega)$, Eqs.(\ref{ha},\ref{hb}) can be recast as
\begin{eqnarray}
\hat{A}_{k}^{(o)} &=& \cosh {(r_k G)} \hat{A}_{k} + \sinh {(r_k G)}\hat{B} ^\dagger _{k},\cr
\hat{B}_{k}^{(o)} &=& \cosh {(r_k G)} \hat{B}_{k} + \sinh {(r_k G)}\hat{A} ^\dagger_{k}, \label{Ak-evl2}
\end{eqnarray}
where $\hat{A}_{k}^{(o)}\equiv \int d\omega \psi_k^{(a)*} \hat a^{(o)}(\omega)$, $\hat{B}_{k}^{(o)}\equiv \int d\omega \psi_k^{(b)*} \hat b^{(o)}(\omega)$, $\hat{A}_{k}\equiv \int d\omega \phi_k^{(a)} \hat a(\omega)$, $\hat{B}_{k}\equiv \int d\omega \phi_k^{(b)}$ $\hat b(\omega)$ define the annihilation operators for the corresponding output and input temporal modes, similar to those given in Eq.(\ref{Ak-evl}). Because of orthonormal relations, they satisfy the Boson commutation relation for annihilation operators.

Because of the $\delta$-function in the initial conditions in Eq.(\ref{ini}), we cannot solve directly the differential-integral equations in Eqs.(\ref{h1a2b},\ref{h1b2a}). In order to proceed, let us write $\bar h_{1a}(\omega_1, \omega_1',z)\equiv h_{1a}(\omega_1, \omega_1',z)-\delta(\omega_1-\omega_1')$ and $\bar h_{1b}(\omega_1, \omega_1',z)\equiv h_{1b}(\omega_1, \omega_1',z)-\delta(\omega_1-\omega_1')$. Furthermore, for a specific parametric process from four-wave mixing in optical fiber with a Gaussian pumping profile, we have $A_p(\omega_3) = A_0^2\exp [-(\omega_3-2\omega_{p0})^2/4\sigma_p^2]$ with pump bandwidth of $\sigma_p$ and pump amplitude $A_0$ \cite{kumar}. We can introduce some new dimensionless variables: $\zeta \equiv z/L_0$, $\Omega_j \equiv (\omega_j-\omega_{j0})/\sigma_p$ ($j=1,2$) with $L_0$ as the length of the nonlinear medium, and $\omega_{j0} (j=1,2,p)$ as the central frequency of the corresponding fields. Then, we can make Eqs.(\ref{h1a2b},\ref{h1b2a}) dimensionless as
\begin{eqnarray}
&&\frac{d}{d\zeta}\bar h_{1a}(\Omega_1, \Omega_1',\zeta)
= \int d\Omega_2  f(\Omega_1, \Omega_2,\zeta)  h_{2b}^*(\Omega_1',\Omega_2,\zeta)\cr
&& \frac{d}{d\zeta}h_{2b}(\Omega_1', \Omega_2, \zeta)
= f(\Omega_1', \Omega_2,\zeta) \cr && \hskip 0.6in + \int d\Omega_1  f(\Omega_1, \Omega_2,\zeta)  \bar h_{1a}^*(\Omega_1,\Omega_1',\zeta),~~~~~~ \label{h1a2b3}
\end{eqnarray}
and
\begin{eqnarray}
&&\frac{d}{d\zeta}\bar h_{1b}(\Omega_2, \Omega_2',\zeta)
=  \int d\Omega_1  f(\Omega_1, \Omega_2,\zeta)  h_{2a}^*(\Omega_1,\Omega_2',\zeta)\cr
&&\frac{d}{d\zeta}h_{2a}(\Omega_1, \Omega_2',\zeta)
= f(\Omega_1, \Omega_2',\zeta) \cr && \hskip 0.6in + \int d\Omega_2  f(\Omega_1, \Omega_2,\zeta)  \bar h_{1b}^*(\Omega_2,\Omega_2',\zeta),~~~~~~ \label{h1b2a3}
\end{eqnarray}
where $f(\Omega_1, \Omega_2,\zeta)\equiv K e^{-i\zeta \Delta k L_0} \exp[-(\Omega_1+\Omega_2)^2/4]$ with $K\equiv 2\pi L_0 A_0^2\sigma_p\chi$ as the dimensionless pump parameter and $h(\Omega_1,\Omega_2)$-functions are dimensionless and is related to $h(\omega)$-functions by $h(\Omega_1,\Omega_2)\equiv \sigma_p h(\omega_1,\omega_2)$. Phase mismatch $\Delta k L_0$ can be adjusted according to the dispersion of the nonlinear medium and in general has a linear form of $\Delta k L_0 = \Omega_1/\Delta_1+\Omega_2/\Delta_2$ with parameters $\Delta_1, \Delta_2$ determined by medium dispersion, pump bandwidth $\sigma_p$, and medium length $L_0$. The initial conditions in Eq.(\ref{ini}) change to
\begin{eqnarray}
\bar h_{1a}(\Omega_1, \Omega_1', \zeta=0) &=& 0\cr
\bar h_{1b}(\Omega_2, \Omega_2', \zeta=0) &=& 0 \cr
h_{2a}(\Omega_1, \Omega_2', \zeta=0) &=& 0\cr
h_{2b}(\Omega_1', \Omega_2, \zeta=0) &=& 0. \label{ini2}
\end{eqnarray}
Note that when pump parameter is small: $K\ll 1$, $h_{2a}(\Omega_1, \Omega_2)$ has an approximate analytical solution:
\begin{eqnarray}
&& h_{2a}(\Omega_1, \Omega_2, \zeta=1) \approx K e^{-i\Delta k L_0/2}{\rm sinc} (\Delta k L_0/2) \cr && \hskip 1.4 in \times\exp[-(\Omega_1+\Omega_2)^2/4],~~~~~~ \label{h2b-0}
\end{eqnarray}
which is exactly the joint spectral function $\Phi(\omega_1,\omega_2)$ in Eq.(\ref{Phi}) after changing to dimensionless quantities. But for a sizable $K$, we cannot solve the differential-integral equations in Eqs.(\ref{h1a2b3},\ref{h1b2a3}) analytically. Next, we will solve them numerically, subject to initial conditions in Eq.(\ref{ini2}).

\begin{figure}
\centering
\includegraphics[width=7.5cm]{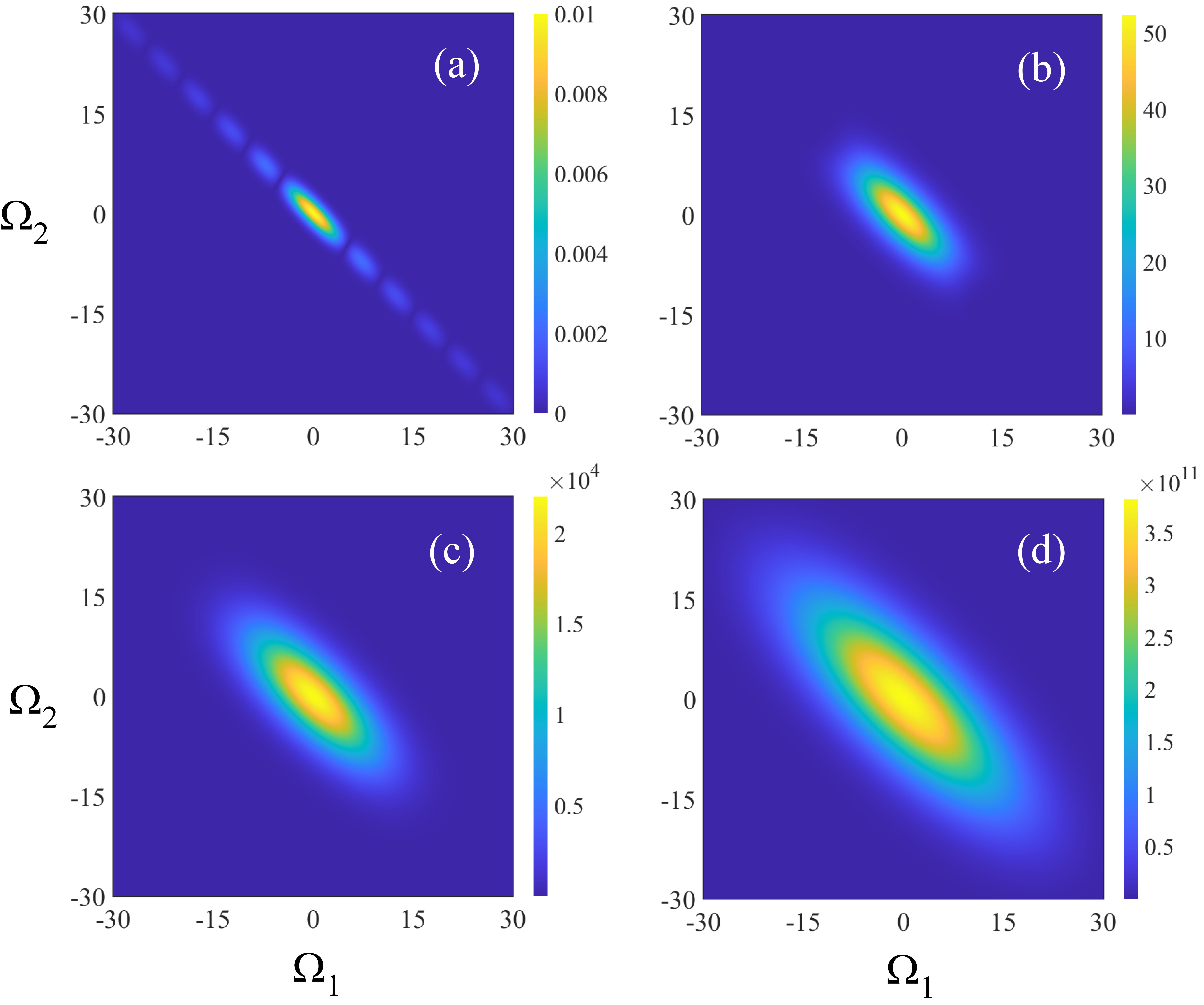}
\caption{Contour plot of the amplitude of $h_{2a}(\Omega_1,\Omega_2)$ for $K=$ (a) 0.01, (b) 2, (c) 4, (d) 10. $\Omega_i \equiv (\omega_i-\omega_{i0})/\sigma_p (i=1,2)$.
}
\label{fig-h2b}
\end{figure}

\section{Numerical Solutions}

\begin{figure}
\centering
\includegraphics[width=7.5cm]{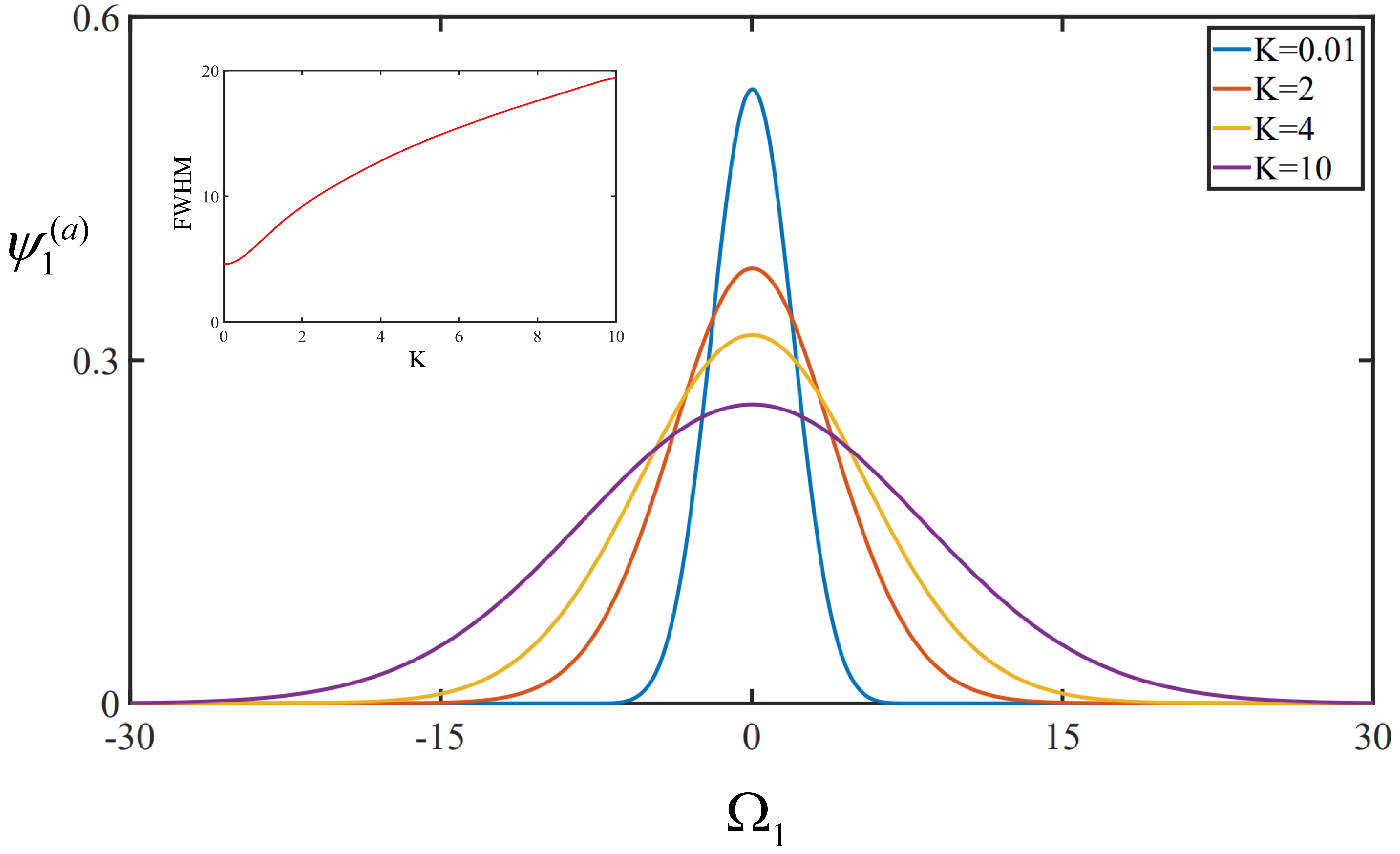}
\caption{ The amplitude function $\psi_1^{(a)}$ of mode 1 for pump parameter $K = 0.01, 2, 4, 10$, showing the broadening of the width. Inset: the full width at half maximum (FWHM) of mode 1 function as a function of the pump parameter $K$. }
\label{fig:mode1}
\end{figure}

Let us use a nonlinear fiber as the nonlinear medium. We use the parameters similar to those given in Ref.\citenum{guo15} for a real piece of 300m-long dispersion-shifted nonlinear fiber. We obtain the dimensionless parameters $1/\Delta_1=0.785, 1/\Delta_2=-0.471$. The numerical solution of the amplitude of $h_{2a}(\Omega_1, \Omega_2)$ is shown in Fig.\ref{fig-h2b} for four values of pump parameter $K$: $K=0.01, 2, 4, 10$. For small $K(\ll 1$, Fig.\ref{fig-h2b}(a)), $h_{2a}(\Omega_1, \Omega_2)$ is exactly the joint spectral function $\Phi(\omega_1,\omega_2)$ given in Eq.(\ref{Phi}), similar to Fig.2(a) of Ref.\citenum{guo15}. The shape starts to broaden as $K$ increases. This can be seen in the profile change of the eigen-function $\psi_1^{(a)}(\Omega_1)$ obtained from singular value decomposition when we plot it in Fig.\ref{fig:mode1} for four values of $K$ (0.01, 2, 4, 10). To show the trend, we plot the full width at half maximum (FWHM) of $|\psi_1^{(a)}(\Omega_1)|$ as a function of $K$ in the inset of Fig.\ref{fig:mode1}.

\begin{figure}
\centering
\includegraphics[width=7.5cm]{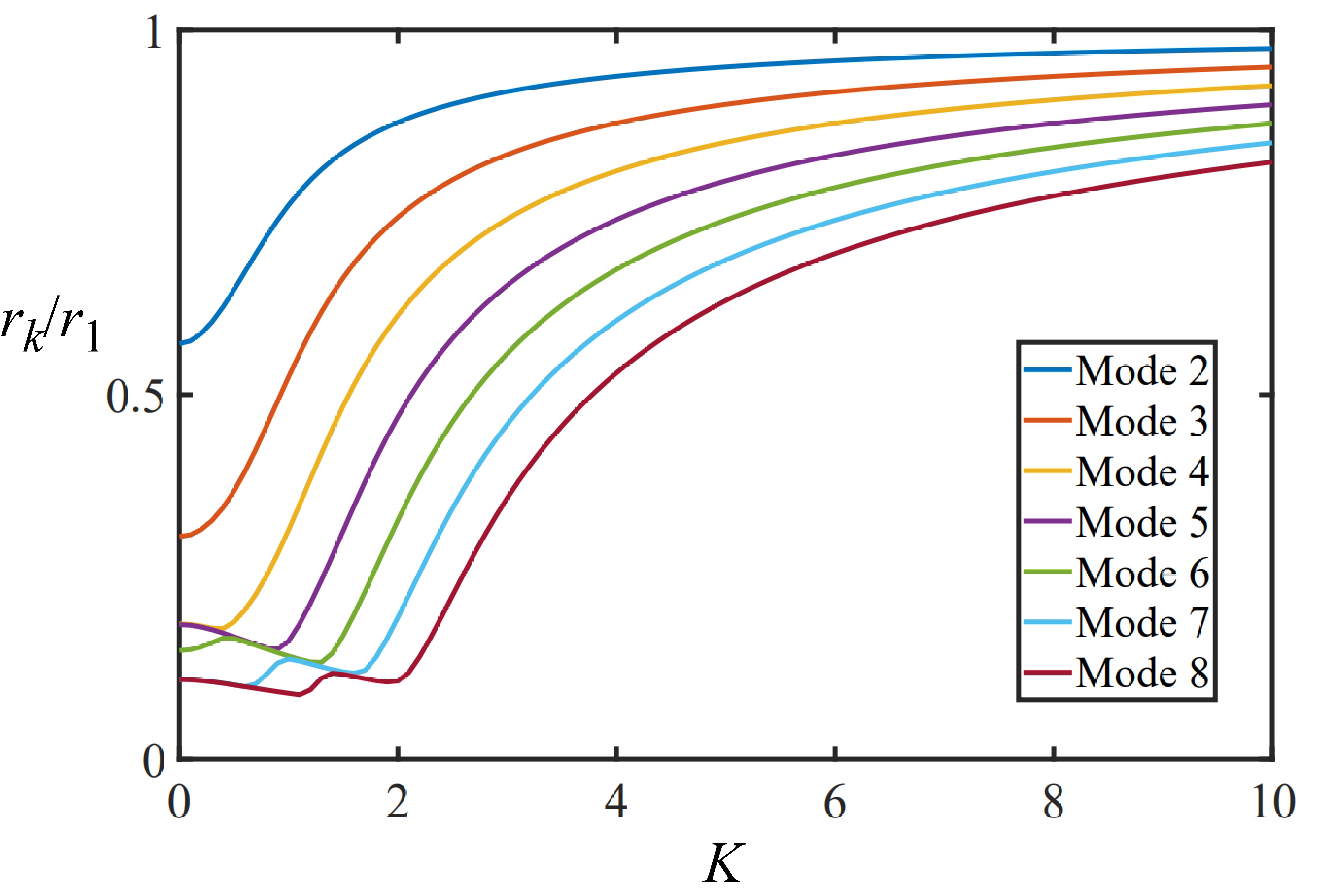}
\caption{Mode coefficients $r_k/r_1$ as a function of dimensionless pump parameter $K$.
}
\label{fig:rk}
\end{figure}

The change of the transfer function $h_{2a}(\Omega_1, \Omega_2)$ with pump parameter $K$ will lead to mode structure change. This is reflected in the change of the distribution of the mode coefficients $\{r_k\}$, whose normalized values to $r_1$ are plotted in Fig.\ref{fig:rk} as a function of the pump parameter $K$. The trend shows the increasing weight of the higher order modes in addition to the broadening of the mode functions as $K$ increases.

\begin{figure}
\centering
\includegraphics[width=7.5cm]{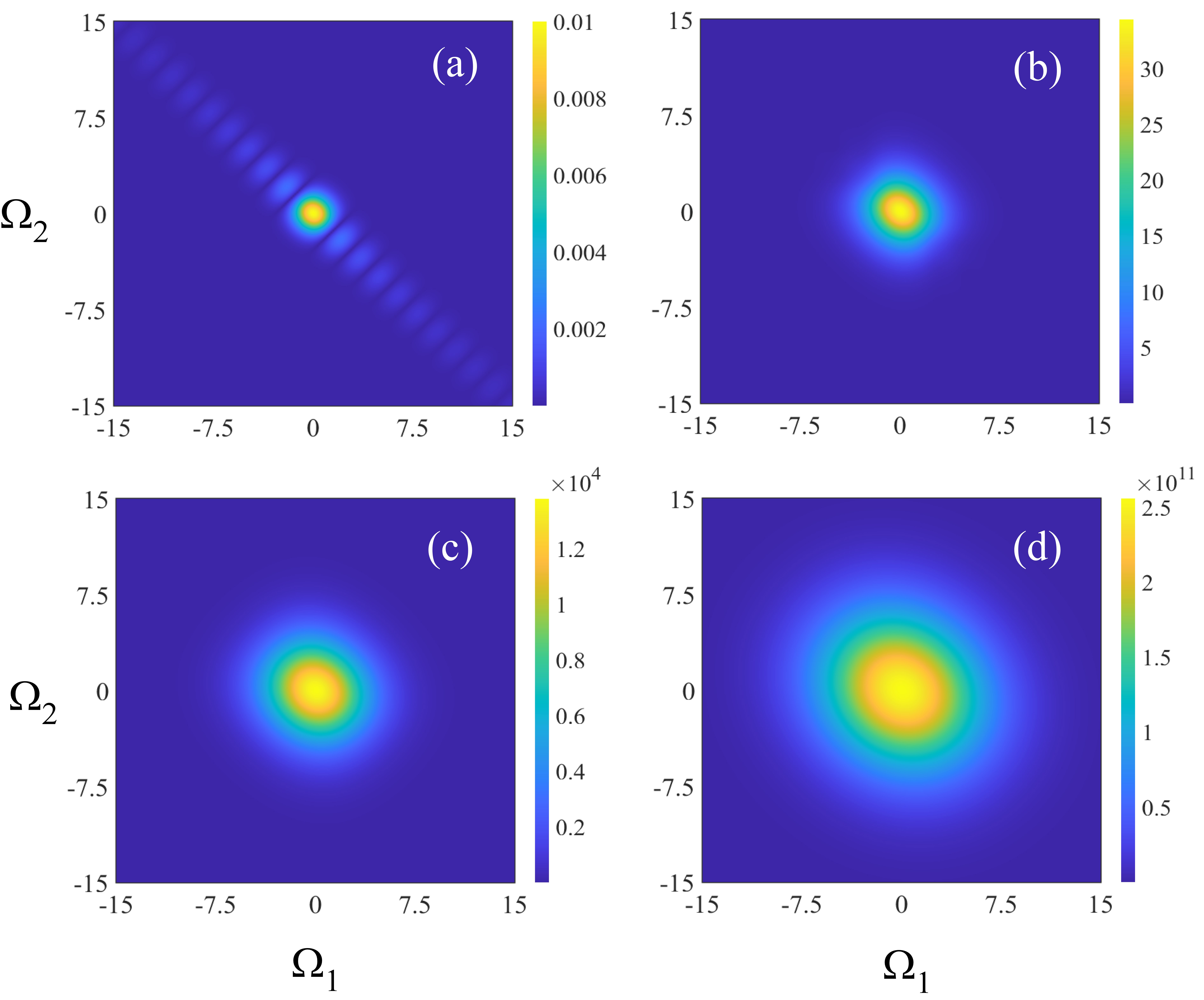}
\caption{Contour plot of the amplitude of $h_{2a}(\Omega_1,\Omega_2)$ for the case of $1/\Delta_1=2.198, 1/\Delta_1=-2.198$ with $K=$ (a) 0.01, (b) 2, (c) 4, (d) 10. $\Omega_i \equiv (\omega_i-\omega_{i0})/\sigma_p (i=1,2)$.
}
\label{fig-h2b-f}
\end{figure}

\begin{figure}
\centering
\includegraphics[width=7.5cm]{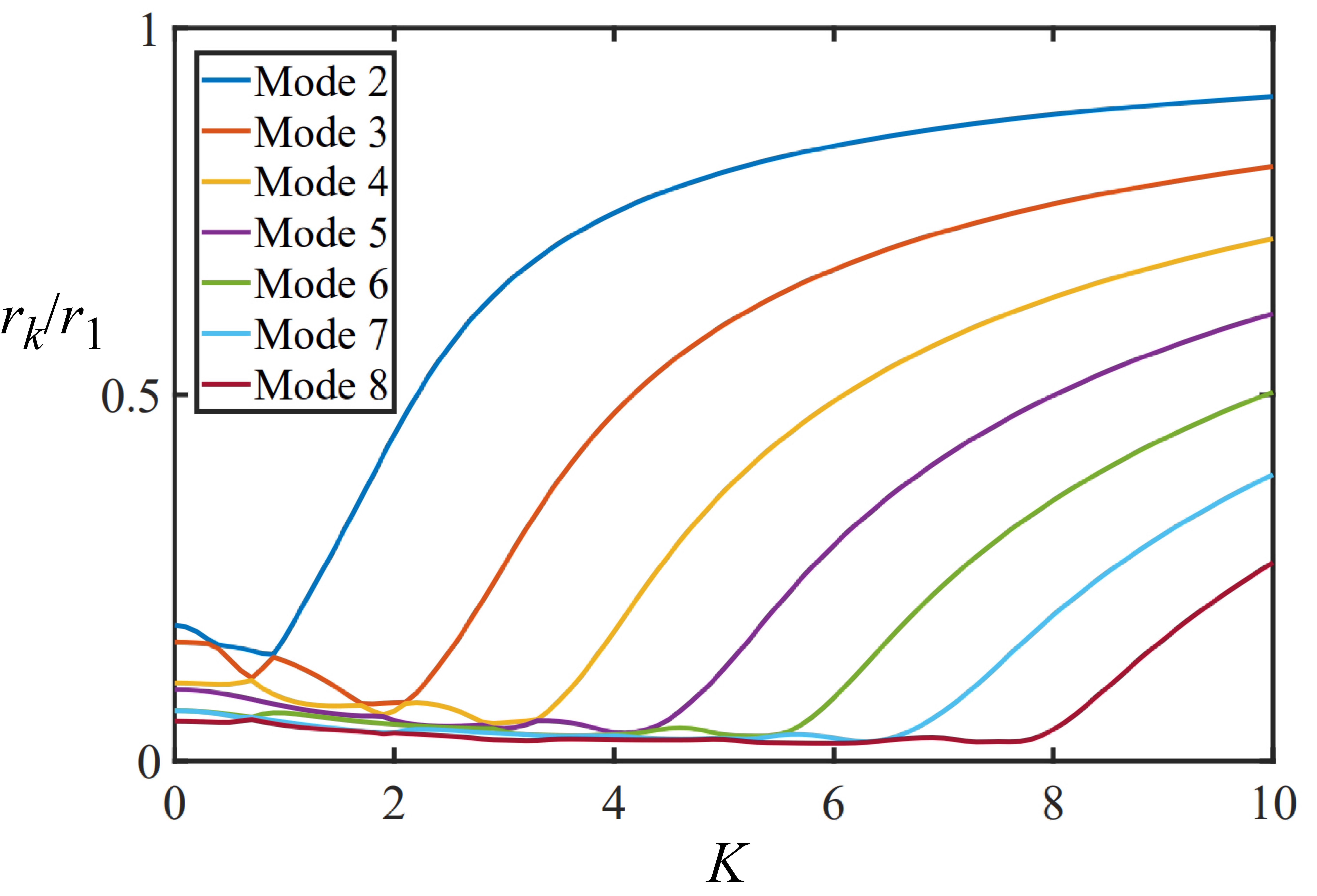}
\caption{Mode coefficients $r_k/r_1$ as a function of dimensionless pump parameter $K$ for the nearly factorized case of $1/\Delta_1=2.198, 1/\Delta_1=-2.198$.
}
\label{fig:rk2}
\end{figure}

The multi-mode nature and the broadening of the mode functions with pump parameter $K$ are due to the values of $\Delta_1,\Delta_2$ in $\Delta kL_0$ for a realistic nonlinear fiber case, which gives rise to an asymmetric sinc-function in $\Phi$ of Eq.(\ref{Phi}) or $h_{2a}(\Omega_1,\Omega_2)$ at small $K$ (Fig.\ref{fig-h2b}(a)). In principle, we can adjust the dispersion parameters of the fiber to change $\Delta_1,\Delta_2$ in $\Delta kL_0$. It is found that  the initial $h_{2a}(\Omega_1,\Omega_2)$ when $K\ll 1$  is nearly round or factorized with parameters $1/\Delta_1=2.198, 1/\Delta_1=-2.198$, as shown in $h_{2a}(\Omega_1,\Omega_2)$ in Fig.\ref{fig-h2b-f}(a). The initial $r_k$ distribution is indeed close to single mode with high order $r_k$ much smaller than 1. This can be seen in Fig.\ref{fig:rk2} for $K=0$. However,
the trend of mode spreading for large $K$ in Fig.\ref{fig:rk} persists in Fig.\ref{fig:rk2}.

On the other hand, a close look at Fig.\ref{fig-h2b-f}(c,d) for large $K$ shows that function $h_{2a}$ is still pretty round or nearly single mode. In fact, we find from mode decomposition in Eq.(\ref{hrk}) that the coefficient of each mode is $\sinh ^2(r_k G)$, which becomes $0.25 e^{2r_kG}$ for large $r_kG$. So, the ratio of coefficients of the first mode to higher mode is then $e^{2(r_1-r_k)G} \gg 1$ for large $G$, that is, the first mode will dominate in the mode decomposition in Eq.(\ref{hrk}) at large $K$ (or $G$) \cite{liunn}, which leads to a nearly round (or factorized) $h_{2a}$.

\begin{figure}
\centering
\includegraphics[width=7cm]{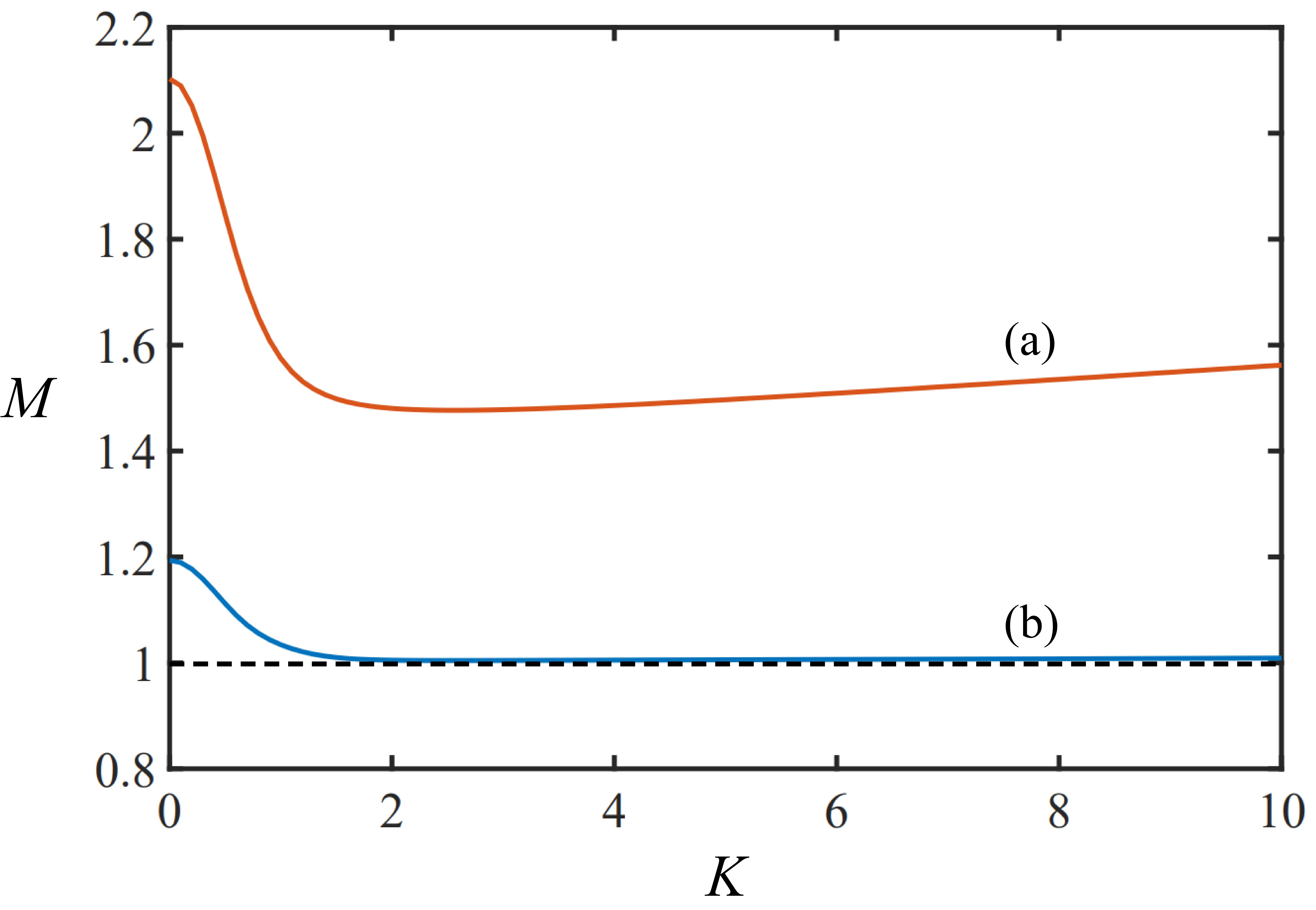}
\caption{Mode number $M$ as a function of pump parameter $K$ for the cases shown in Fig.\ref{fig-h2b} (orange, (a)) and Fig.\ref{fig-h2b-f} (blue, (b)). The dashed line corresponds to $M=1$.
}
\label{fig-M}
\end{figure}

Furthermore, we can look at the mode purity of the output state. It is known that if we discard the other field, one of the output fields from a parametric amplifier is in a thermal state. For broadband pulsed pumping, it becomes a multi-mode thermal field.  Mode purity for a pulsed multi-mode thermal field was studied in Ref.\citenum{su19PRA} with the temporal mode format described here.  It was found that mode purity can be characterized by the normalized intensity correlation function $g^{(2)}$ defined as
\begin{eqnarray}
g^{(2)} & \equiv & \frac{\langle I^2\rangle}{\langle I\rangle^2} \equiv 1 + \frac{1}{M} \cr &=& 1 + \frac{\sum_k I_k^2}{(\sum_k I_k)^2}, \label{g2}
\end{eqnarray}
where $M$ is defined as the number of modes ($M=1$ gives the pure single-mode case) and $I_k$ is the intensity of mode $k$. From Eq.(\ref{Ak-evl2}), we find that $I_k=\sinh^2(r_kG)$ for vacuum input to the parametric amplifier. So, Eq.(\ref{g2}) becomes
\begin{eqnarray}
M =  \frac{\big[\sum_k \sinh^2(r_kG)\big]^2}{\sum_k \sinh^4(r_kG)}. \label{M}
\end{eqnarray}
Obviously, for the single mode case with $r_1=1, r_k=0~(k > 1)$, we have $M=1$. Equation (\ref{M}) was first derived by Christ {\it et al.} \cite{sil} for temporal modes and by Sharapova {\it et al.} \cite{sha} and Dyakonov {\it et al.} \cite{dya} for spatial modes in high gain parametric processes. For the two cases shown in Fig.\ref{fig-h2b} and Fig.\ref{fig-h2b-f}, we evaluate mode number $M$ in Eq.(\ref{M}) as a function of pump parameter $K$ and plot the results in Fig.\ref{fig-M}. It can be seen that mode number $M$ indeed drops first as $K$ increases. This drop is consistent with the dominance of the first mode as $K$ increases. However, the drop stops at around $K=2$ when a minimum of $M$ is reached. $M$ starts to slowly increase after $K> 2$. This is due to the increase of $r_k$ for higher order modes as $K$ increases, as shown in Fig.\ref{fig:rk} and Fig.\ref{fig:rk2}. The minimum $M$ value depends on the initial $M$ at low $K(\ll 1)$. Thus, it is better to have a nearly single mode situation at low pump power.

\section{Summary and Discussion}

We studied the mode structure for a broadband parametric amplifier at different gains by using an input-output approach that avoids the crucial issue of time-ordering in Hamiltonian. Contrary to previous studies where the time-ordering issue was not treated, the mode structure changes as the gain increases in the sense that both the mode distribution and the mode functions broaden with the increase of the gain. Although the mode number, a quantity that characterizes the total number of modes, drops initially as the gain changes from low to high due to the dominance of the first mode, it reaches a minimum value before slowly increases due to the broadening of the mode distribution.

The mode structure change with the gain will have profound impact on the application of broadband parametric processes in quantum technology with continuous variables, which relies on homodyne detection method as the dominant measurement technique. Even though the output of the broadband parametric processes is of multi-mode nature, because of the orthogonality of the modes in the processes, we can choose the local oscillators of homodyne measurement to match the mode functions of the output and select specific modes for study \cite{huo20}. The change of mode structure and functions with the gain means that we will need to measure them constantly at different gains as we change the operation condition. Fortunately, methods are recently developed for the direct measurement of mode functions and mode coefficients of parametric processes \cite{huo20,chen20}.

\begin{acknowledgments}
This work was supported  by US National Science Foundation (Grant No. 1806425).
\end{acknowledgments}

\section{appendix}
In general, we can use singular value decomposition to write
\begin{eqnarray}\label{svd}
&&h_{1a}(\omega,\omega') = \sum_{k}r_{1a}^{(k)}\psi_{1a}^{(k)}(\omega)\phi_{1a}^{(k)}(\omega') \cr
&&h_{2a}(\omega,\omega') =\sum_{k}r_{2a}^{(k)}\psi_{2a}^{(k)}(\omega)\phi_{2a}^{(k)}(\omega') \cr
&&h_{1b}(\omega,\omega') =\sum_{k}r_{1b}^{(k)}\psi_{1b}^{(k)}(\omega)\phi_{1b}^{(k)}(\omega') \cr
&&h_{2b}(\omega,\omega') = \sum_{k}r_{2b}^{(k)}\psi_{2b}^{(k)}(\omega)\phi_{2b}^{(k)}(\omega')
\end{eqnarray}
where $\psi_{1a}^{(k)}(\omega), \phi_{1a}^{(k)}(\omega')$ etc. are 8 sets of orthonormal mode functions satisfying $\int d\omega \psi_{1a}^{(k)}(\omega)$ $\psi_{1a}^{(k')*}(\omega) = \delta_{k,k'}$, etc. Using the the decomposition in Eq.(\ref{svd}) and orthonormal relations, we write the left hand side of the first relation in Eq.(\ref{h-rel}) as
\begin{eqnarray}
&&\int d\omega''h_{1a}(\omega,\omega'')h_{1a}^*(\omega',\omega'') \cr && \hskip 0.8in -\int d\omega''h_{2a}(\omega,\omega'')h_{2a}^*(\omega',\omega'')
\cr
&&=\int d\omega''\sum_{k,k'}r_{1a}^{(k)}r_{1a}^{(k')} \psi_{1a}^{(k)}(\omega)\psi_{1a}^{(k')*}(\omega')\phi_{1a}^{(k)}(\omega'')\phi_{1a}^{(k')*}(\omega'')
\cr &&
- \int d\omega''\sum_{k,k'}r_{2a}^{(k)}r_{2a}^{(k')} \psi_{2a}^{(k)}(\omega)\psi_{2a}^{(k')*}(\omega')\phi_{2a}^{(k)}(\omega'')\phi_{2a}^{(k')*}(\omega'')
\cr &&
=\sum_{k,k'}r_{1a}^{(k)}r_{1a}^{(k')}\psi_{1a}^{(k)}(\omega)\psi_{1a}^{(k')*}(\omega')\delta_{k,k'}
\cr && \hskip 0.6in
 - \sum_{k,k'}r_{2a}^{(k)}r_{2a}^{(k')}\psi_{2a}^{(k)}(\omega)\psi_{2a}^{(k')*}(\omega')\delta_{k,k'}
\cr &&
=\sum_{k}r_{1a}^{(k)2}\psi_{1a}^{(k)}(\omega)\psi_{1a}^{(k)*}(\omega') - \sum_{k}r_{2a}^{(k)2}\psi_{2a}^{(k)}(\omega)\psi_{2a}^{(k)*}(\omega').\cr &&
\end{eqnarray}
Using the completeness of mode function $\{\psi_{1a}^{(k)}(\omega)\}$:
$\sum\limits_{k}\psi_{1a}^{(k)}(\omega)\psi_{1a}^{(k)*}(\omega') =\delta(\omega-\omega')$,
we can rewrite the first relation in Eq.(\ref{h-rel}) as
\begin{eqnarray}\label{sum-rk}
&&\sum_{k}r_{2a}^{(k)2}\psi_{2a}^{(k)}(\omega)\psi_{2a}^{(k)*}(\omega')\cr && \hskip 0.6in =
\sum_{k}(r_{1a}^{(k)2}-1)\psi_{1a}^{(k)}(\omega)\psi_{1a}^{(k)*}(\omega').
\end{eqnarray}

Treating $k$ as the row index and $\omega$ as the column index of a matrix, we can consider mode function $\psi^{(k)}(\omega)$ as matrix element of $\Psi$ with $\{\Psi\}_{k,\omega} \equiv \psi^{(k)}(\omega)$ (we dropped subscript $1a,2a$ for clarity). Then Eq.(\ref{sum-rk}) is equivalent to the following matrix equation:
\begin{eqnarray}
&& [\Psi_{2a}]^{\dag}
\begin{pmatrix}
r_{2a}^{(1)2} &0 & \dots\cr
�0& r_{2a}^{(2)2} &\dots\cr
�\vdots &\vdots & \ddots
\end{pmatrix}
\Psi_{2a} \cr && \hskip 0.3in = [\Psi_{1a}]^{\dag}
\begin{pmatrix}
r_{1a}^{(1)2}-1 &0 & \dots \cr
�0& r_{1a}^{(2)2}-1 &\dots \cr
�\vdots &\vdots & \ddots
\end{pmatrix} \Psi_{1a}.
\end{eqnarray}
Multiplying left of the above with matrix $\Psi_{1a}$ and right with $[\Psi_{1a}]^{\dag}$ and using orthonormal relation $\int d\omega \psi_{1a}^{(k)}(\omega) \psi_{1a}^{(k')*}(\omega) = \delta_{k,k'}$, we obtain a rotational transformation:
\begin{eqnarray}
&&{\rm R}
\begin{pmatrix}
r_{2a}^{(1)2} &0 & \dots \cr
�0& r_{2a}^{(2)2} &\dots \cr
�\vdots &\vdots & \ddots
\end{pmatrix}
{\rm R}^{\dag}\cr && \hskip 0.6in=
\begin{pmatrix}
r_{1a}^{(1)2}-1 &0 & \dots \cr
�0& r_{1a}^{(2)2}-1 &\dots \cr
�\vdots &\vdots & \ddots
\end{pmatrix}
\end{eqnarray}
with transformation matrix ${\rm R} \equiv \Psi_{1a} [\Psi_{2a}]^{\dag}$.
The only solution for the above is $\rm R=I$, or $\int d\omega \psi_{1a}^{(k)}(\omega)\psi_{2a}^{(k')*}(\omega) = \delta_{k,k'}$.
With uniqueness of mode function set $\{\psi_{1a}^{(k)}(\omega)\}$, we have
$
\psi_{1a}^{(k)}(\omega)=\psi_{2a}^{(k)}(\omega),
$
and then
$
r_{1a}^{(k)2}-r_{2a}^{(k)2}=1
$.
Similarly, using the second relation in Eq.(\ref{h-rel}) we have
$
\psi_{1b}^{(k)}(\omega)=\psi_{2b}^{(k)}(\omega),
$
and
$
r_{1b}^{(k)2}-r_{2b}^{(k)2}=1.
$

Next let us use the third relation in Eq.(\ref{h-rel}) and rewrite as:
\begin{eqnarray}
&&\int d\omega''\sum_{k,k'}r_{1a}^{(k)}r_{2b}^{(k')} \psi_{1a}^{(k)}(\omega)\psi_{1b}^{(k')}(\omega')\phi_{1a}^{(k)}(\omega'')\phi_{2b}^{(k')}(
\omega'')
\cr && \hskip 0.6in
- \int d\omega''\sum_{k,k'}r_{2a}^{(k)}r_{1b}^{(k')} \psi_{1a}^{(k)}(\omega)\psi_{1b}^{(k')}(\omega')\cr && \hskip 1.3in \times \phi_{2a}^{(k)}(\omega'')\phi_{1b}^{(k')}(
\omega'')
= 0.
\end{eqnarray}
Multiplying both sides with $\psi_{1a}^{(k_1)*}(\omega)\psi_{1b}^{(k_2)*}(\omega')$ and integrating $\omega,\omega'$, with orthonormal relations for $\psi_{1a}^{(k)}(\omega), \psi_{1b}^{(k')}(\omega')$,
we obtain
\begin{eqnarray}\label{kk'2}
&&\cosh{r_a^{(k_1)}}\sinh{r_b^{(k_2)}}R_{k_2,k_1}\cr && \hskip 0.8in =\sinh r_a^{(k_1)}\cosh{r_b^{(k_2)}}R_{k_2,k_1}',
\end{eqnarray}
where we set
$ r_{1a}^{(k)} \equiv \cosh{r_a^{(k)}}
$, $r_{1b}^{(k)} \equiv \cosh{r_b^{(k)}}$, then
$r_{2a}^{(k)} = \sqrt{r_{1a}^{(k)2}-1} = \sinh{r_a^{(k)}}
$
and
$r_{2b}^{(k)} =\sqrt{r_{1b}^{(k)2}-1} = \sinh{r_b^{(k)}}$, and $R_{k_2,k_1}' \equiv \int d\omega \phi_{1b}^{(k_2)}(\omega)\phi_{2a}^{(k_1)}(\omega)$, $R_{k_2,k_1} \equiv \int d\omega \phi_{2b}^{(k_2)}(\omega)\phi_{1a}^{(k_1)}(\omega)$.  Switching back notations: $k=k_1, k'=k_2$, we have
\begin{eqnarray}\label{kk'3}
R_{k',k}&=&\tanh{r_a^{(k)}}\coth{r_b^{(k')}}R_{k',k}',\cr R_{k',k}'&=&\coth{r_a^{(k)}}\tanh{r_b^{(k')}}R_{k',k},
\end{eqnarray}
which, in matrix form, is simply ${\rm R} = {\rm C}_b{\rm R}'{\rm C}_a^{-1}$ or ${\rm R}' = {\rm C}_b^{-1}{\rm R}{\rm C}_a$ with
\begin{eqnarray}
{\rm C}_a \equiv
\begin{pmatrix}
\coth{r_a^{(1)}} &0 & \dots \cr
�0& \coth{r_a^{(2)}} &\dots \cr
\vdots & \vdots & \ddots
\end{pmatrix}
\end{eqnarray}
and
\begin{eqnarray}
{\rm C}_b \equiv
\begin{pmatrix}
\coth{r_b^{(1)}} &0 & \dots \cr
�0& \coth{r_b^{(2)}} &\dots \cr
\vdots &\vdots & \ddots
\end{pmatrix}.
\end{eqnarray}
Now notice that ${\rm RR^{\dag}= I = R'R'^{\dag}}$, or using matrix form of Eq.(\ref{kk'3}), we obtain
\begin{eqnarray}
{\rm C}_b{\rm R'C}_a^{-1}{\rm C}_a^{-1}{\rm R'^{\dag}C}_b={\rm I}, ~
{\rm C}_b^{-1}{\rm R C}_a{\rm C}_a{\rm R^{\dag}C}_b^{-1}={\rm  I}.~~~~
\end{eqnarray}
Rewrite the above, we have
\begin{eqnarray}
{\rm R'C}_a^{-2}{\rm R'^{\dag}}= {\rm C}_b^{-2} , ~~
{\rm R C}_a^{2}{\rm R^{\dag}}= {\rm C}_b^{2}.
\end{eqnarray}
Since matrices ${\rm C}_a, {\rm C}_b$ are both diagonalized, the above expressions are true only if ${\rm R' = I}$ $= {\rm R}$, or $\phi_{1a}^{(k)}=\phi_{2b}^{(k)*}$, $\phi_{1b}^{(k)}=\phi_{2a}^{(k)*}$ and
${\rm C}_a= {\rm C}_b$, or $r_a^{(k)}=r_b^{(k)}$.

Setting $r_a^{(k)}=r_b^{(k)} \equiv r_k G$ with $r_k$ normalized: $\sum_k r_k^2$ $=1$, and $\phi_{1a}^{(k)}=\phi_{2b}^{(k)*}\equiv \phi_{k}^{(a)}, \phi_{1b}^{(k)}=\phi_{2a}^{(k)*} \equiv \phi_{k}^{(b)},
\psi_{1a}^{(k)}(\omega)$ $ =\psi_{2a}^{(k)}(\omega)\equiv \psi_{k}^{(a)}(\omega), \psi_{1b}^{(k)}(\omega)=\psi_{2b}^{(k)}(\omega)\equiv \psi_{k}^{(b)}(\omega)
$ in Eq.(\ref{svd}), we obtain Eq.(\ref{hrk}).




\begin{thebibliography}{10}
\newcommand{\enquote}[1]{``#1''}

\bibitem{walls} D. F. Walls and G. J. Milburn, {\it Quantum Optics}, 2nd ed. Springer (2008).

\bibitem{reid} M. D. Reid, P. D. Drummond, W. P. Bowen, E. G. Cavalcanti, P. K. Lam, H. A. Bachor, U. L. Andersen, and G. Leuchs, \enquote{The Einstein-Podolsky-Rosen paradox: From concepts to applications,} Rev. Mod. Phys. {\bf 81}, 1727 (2009).

\bibitem{kumar} Orhan Ayt\"ur and Prem Kumar, \enquote{Pulsed twin beams of light,} Phys. Rev. Lett. {\bf 65}, 1551 (1990).

\bibitem{wg} A. Eckstein, A. Christ, P. J. Mosley, and C. Silberhorn, \enquote{Highly Efficient Single-Pass Source of Pulsed Single-Mode Twin Beams of Light,} Phys. Rev. Lett. {\bf 106}, 013603 (2011).

\bibitem{guo16} Xueshi Guo, Nannan Liu, Yuhong Liu, Xiaoying Li, and Z. Y. Ou, \enquote{Generation of continuous variable quantum entanglement using a fiber optical parametric amplifier,} Opt. Lett. {\bf 41}, 653 (2016).

\bibitem{guo13} Xueshi Guo, Xiaoying Li, Nannan Liu, and Z. Y. Ou, \enquote{Multimode theory of pulsed-twin-beam generation using a high-gain fiber-optical parametric amplifier,} Phys. Rev. A {\bf 88}, 023841 (2013).

\bibitem{lvo} W. Wasilewski, A. I. Lvovsky, K. Banaszek,1 and C. Radzewicz, \enquote{Pulsed squeezed light: Simultaneous squeezing of multiple modes,} Phys. Rev. A {\bf 73}, 063819 (2006).

\bibitem{sil} A. Christ, K. Laiho, A. Eckstein, K. N. Cassemiro, and C. Silberhorn, \enquote{Probing multimode squeezing with correlation functions,} New J. Phys. {\bf 13}, 033027 (2011).

\bibitem{law} C. K. Law, I. A. Walmsley, and J. H. Eberly, \enquote{Continuous Frequency Entanglement: Effective Finite Hilbert Space and Entropy Control,} Phys. Rev. Lett. {\bf 84}, 5304 (2000).

\bibitem{cui20} Liang Cui, Jie Su, Jiamin Li, Yuhong Liu, Xiaoying Li, and Z. Y. Ou, \enquote{Quantum state engineering by nonlinear quantum interference,} Phys. Rev. A {\bf 102}, 033718 (2020).

\bibitem{huo20} Nan Huo, Yuhong Liu, Jiamin Li, Liang Cui, Xin Chen, Rithwik Palivela, Tianqi Xie, Xiaoying Li, and Z. Y. Ou, \enquote{Direct Temporal Mode Measurement for the Characterization of Temporally Multiplexed High Dimensional Quantum Entanglement in Continuous Variables,} Phys. Rev. Lett. {\bf 124}, 213603 (2020).

\bibitem{sipe} N. Quesada and J. E. Sipe, \enquote{Effects of time ordering in quantum
nonlinear optics,} Phys. Rev. A {\bf 90}, 063840 (2014).

\bibitem{sha20} P. R. Sharapova, G. Frascella, M. Riabinin, A. M. P\'erez, O. V. Tikhonova, S. Lemieux, R. W. Boyd, G. Leuchs, and M. V. Chekhova, \enquote{Properties of bright squeezed vacuum at increasing brightness,} Phys. Rev. Research {\bf 2}, 013371 (2020).

\bibitem{sil11} A. Eckstein, B. Brecht, and C. Silberhorn, \enquote{A quantum pulse
gate based on spectrally engineered sum frequency generation,}
Opt. Express {\bf 19}, 13770 (2011).

\bibitem{sil14} B. Brecht, A. Eckstein, R. Ricken, V. Quiring, H. Suche,
L. Sansoni, and C. Silberhorn, \enquote{Demonstration of coherent
time-frequency schmidt mode selection using dispersionengineered
frequency conversion,} Phys. Rev. A {\bf 90}, 030302(R)
(2014).

\bibitem{ray14} D. V. Reddy, M. G. Raymer, and C. J. McKinstrie, \enquote{Efficient
sorting of quantum-optical wave packets by temporal-mode
interferometry,} Opt. Lett. {\bf 39}, 2924 (2014).

\bibitem{ou12} Z. Y. Ou, \enquote{Enhancement of the phase-measurement sensitivity beyond the standard quantum limit by a nonlinear interferometer,} Phys. Rev. A {\bf 85}, 023815 (2012).

\bibitem{ouli20} Z. Y. Ou and Xiaoying Li, \enquote{Quantum SU(1,1) interferometers: Basic principles and applications,} APL Photonics {\bf 5}, 080902 (2020).

\bibitem{hud14} F. Hudelist, J. Kong, C. Liu, J. Jing, Z. Y. Ou, and W. Zhang, \enquote{Quantum metrology with parametric amplifierbased photon correlation interferometers,} Nat. Commun. {\bf 5}, 3049 (2014).

\bibitem{zei14} G. B. Lemos, V. Borish, G. D. Cole, S. Ramelow, R. Lapkiewicz, and A. Zeilinger, \enquote{Quantum imaging with undetected photons,} Nature {\bf 512}, 409 (2014).

\bibitem{su19} Jie Su, Liang Cui, Jiamin Li, Yuhong Liu, Xiaoying Li, and Z. Y. Ou, \enquote{Versatile and precise quantum state engineering by using nonlinear interferometers}, Opt. Exp. \textbf{27}, 20479 (2019).

\bibitem{li20} Jiamin Li,  Jie Su,  Liang Cui, Tianqi Xie,  Z. Y. Ou, and Xiaoying Li, \enquote{Generation of pure-state single photons with high heralding efficiency by using a three-stage nonlinear interferometer,} Appl. Phys. Lett. {\bf 116}, 204002 (2020).

\bibitem{que20} Nicol\'as Quesada, Gil Triginer, Mihai D. Vidrighin, and J. E. Sipe, \enquote{Theory of high-gain twin-beam generation in waveguides: From Maxwell's equations to efficient simulation,}
Phys. Rev. A {\bf 102}, 033519 (2020).

\bibitem{guo15} Xueshi Guo, Nannan Liu, Xiaoying Li, and Z. Y. Ou, \enquote{Complete temporal mode analysis in pulse-pumped fiberoptical parametric amplifier for continuous variable entanglement generation,} Opt. Express {\bf 23}, 29369 (2015).

\bibitem{liunn} Nannan Liu, Yuhong Liu, Xueshi Guo, Lei Yang, Xiaoying Li, and Z. Y. Ou, \enquote{Approaching single temporal mode operation in twin beams generated by pulse pumped high gain spontaneous four wave mixing,} Opt. Expr. {\bf 24}, 1096 (2016).

\bibitem{su19PRA} Jie Su, Jiaming Li, Liang Cui, Xiaoying Li, and Z. Y. Ou, \enquote{Interference between two independent multi-temporal-mode thermal fields,} Phys. Rev. A {\bf 99}, 013838 (2019).

\bibitem{sha} P. Sharapova, A. M. P\'erez, O. V. Tikhonova, and M. V. Chekhova, \enquote{Schmidt modes in the angular spectrum of bright squeezed vacuum,} Phys. Rev. A {\bf 91}, 043816 (2015).

\bibitem{dya} I. V. Dyakonov, P. R. Sharapova, T. S. Iskhakov, and G. Leuchs, \enquote{Direct Schmidt number measurement of highgain parametric down conversion,} Laser Phys. Lett. {\bf 12}, 065202 (2015).






\bibitem{ou17} Z. Y. Ou, {\it Quantum Optics for Experimentalists} (World Scientific, 2017).

\bibitem{chen20} Xin Chen, Xiaoying Li, and Z. Y. Ou, \enquote{Direct temporal mode measurement of photon pairs by stimulated emission,} Phys. Rev. A {\bf 101}, 033838 (2020).




\end{thebibliography}

\end{document}